\documentclass[%
 reprint,
amsmath,
amssymb,
longbibliography,
prl,
raggedbottom,
lengthcheck,
]{revtex4}

\usepackage{graphicx}% Include figure files
\usepackage{dcolumn}% Align table columns on decimal point
\usepackage{bm}% bold math

\usepackage{mathrsfs}   %for mathscr -- font used for spaces
\usepackage{float}
\usepackage{hyperref}
\usepackage{rotating}
\usepackage{array}
\usepackage{subfigure}
\usepackage{multirow}
\usepackage{xspace}
\usepackage{xifthen}
\usepackage{xfrac}
\usepackage{longtable}
\usepackage{algorithm}
\usepackage{algorithmic}
\usepackage{upgreek}  %upright greek characters
\usepackage{stmaryrd} %for llbracket and rrbracket
\usepackage{framed}
\usepackage{appendix}
\usepackage{tabularx}
\usepackage{xifthen}
\usepackage{setspace}

\newcommand{\mymath}[1]{\ensuremath{#1}}

\newcommand{\TensorOneRelayGreek}[1][]{%
  \ifthenelse{\isempty{#1}}%
    {\mymath{\bs{\ArgI}}}%standard notation, no indices
    {\mymath{\ArgI_{#1}}}%index notation with indices as #1
}
\newcommand{\TensorTwoGreek}[1][T]{%
  \def\ArgI{{#1}}%
  \TensorTwoRelayGreek
}
\newcommand{\TensorTwoRelayGreek}[1][]{%
  \ifthenelse{\isempty{#1}}%
    {\mymath{\bs{\ArgI}}}%standard notation, no indices
    {\mymath{\ArgI_{#1}}}%index notation with indices as #1
}
\newcommand{\TensorThreeRelayGreek}[1][]{%
  \ifthenelse{\isempty{#1}}%
    {\mymath{\bs{\ArgI}}}%standard notation, no indices
    {\mymath{\ArgI_{#1}}}%index notation with indices as #1
}

\newcommand{\TensorOneNonGreek}[1][T]{%
  \def\ArgI{{#1}}%
  \TensorOneRelayNonGreek
}
\newcommand{\TensorOneRelayNonGreek}[1][]{%
  \ifthenelse{\isempty{#1}}%
    {\mymath{{\mathbf \ArgI}}}%standard notation, no indices
    {\mymath{\ArgI_{#1}}}%index notation with indices as #1
}
\newcommand{\TensorTwoNonGreek}[1][T]{%
  \def\ArgI{{#1}}%
  \TensorTwoRelayNonGreek
}
\newcommand{\TensorTwoRelayNonGreek}[1][]{%
  \ifthenelse{\isempty{#1}}%
    {\mymath{{\mathbf \ArgI}}}%standard notation, no indices
    {\mymath{\ArgI_{#1}}}%index notation with indices as #1
}
\newcommand{\TensorFourRelayNonGreek}[1][]{%
  \ifthenelse{\isempty{#1}}%
    {\mymath{\bm{\ArgI}}}%standard notation, no indices
    {\mymath{{\mathit\ArgI}_{#1}}}%index notation with indices as #1
}
\newcommand{\TensorSixRelayNonGreek}[1][]{%
  \ifthenelse{\isempty{#1}}%
    {\mymath{\pmb{\mathsf{\ArgI}}}}%standard notation, no indices
    {\mymath{{\mathsf\ArgI}_{#1}}}%index notation with indices as #1
}

\newcommand{\mathspace}[1]{\mymath{\mathscr{#1}}}

\newcommand{\elastic}{\mymath{\mathrm{e}}}
\newcommand{\plastic}{\mymath{\mathrm{p}}}
\newcommand{\chemistry}{\mymath{\mathrm{c}}}
\newcommand{\thermal}{\mymath{\uptheta}}
\newcommand{\trial}{\mymath{\mathrm{trial}}}

\newcommand{\T}{\mymath{\theta}}

\newcommand{\Tdot}{\mymath{\dot{\theta}}}
\newcommand{\y}{\mymath{y}}

\newcommand{\ydot}{\dot{\y}}

\newcommand{\Mol}{\mymath{M}}
\newcommand{\R}[1][]{r}
\newcommand{\Nsp}{\mymath{N_\mathrm{s}}}
\newcommand{\bd}{\mymath{\beta\rightarrow\delta}}
\newcommand{\dA}{\mymath{\widehat{A}}}
\newcommand{\kc}{\mymath{k_\mathrm{c}}}

\newcommand{\pressure}{\mymath{p}}

\newcommand{\entropy}{\mymath{\eta}}

\newcommand{\enthalpy}{\mymath{h}}

\newcommand{\chempot}{\mymath{\chi}}

\newcommand{\hreact}[1][]{\mymath{\Delta \enthalpy^\mathrm{rxn}_{#1}}}
\newcommand{\ereact}[1][]{\mymath{\Delta \ien^\mathrm{rxn}_{#1}}}

\newcommand{\qheat}[1][]{\vec{\TensorOneNonGreek[q][#1]}}

\newcommand{\qsource}{\mymath{r}}

\newcommand{\Q}[1][]{\mymath{Q}}
\newcommand{\Qe}[1][]{\mymath{Q_\elastic}}
\newcommand{\Qp}[1][]{\mymath{Q_\plastic}}
\newcommand{\Qm}[1][]{\mymath{\mathscr{H}}}
\newcommand{\Qw}[1][]{\mymath{\,^\upomega Q}}
\newcommand{\Qc}[1][]{\mymath{Q_\chemistry}}
\newcommand{\Qs}[1][]{\mymath{\;^\mathrm{s}Q}}

\newcommand{\q}[1][]{\mymath{q}}
\newcommand{\qe}[1][]{\mymath{q_\elastic}}
\newcommand{\qp}[1][]{\mymath{q_\plastic}}
\newcommand{\qw}[1][]{\mymath{\,^\upomega q}}
\newcommand{\qc}[1][]{\mymath{q_\chemistry}}
\newcommand{\qs}[1][]{\mymath{\;^\mathrm{s}q}}

\newcommand{\bforce}[1][]{\vec{\TensorOneNonGreek[f][#1]}}

\newcommand{\Diss}{\mymath{\mathspace{D}}}
\newcommand{\tDiss}{\mymath{\Diss_\thermal}}

\newcommand{\cDiss}{\mymath{\Diss_\chemistry}}

\newcommand{\diff}[1]{\mymath{\;\mathrm{d}{#1}}}

\newcommand{\EyeTwo}[1][]{\TensorTwoNonGreek[1][#1]}
\newcommand{\cte}{\mymath{\alpha}}
\newcommand{\ctek}{\mymath{\omega}}

\newcommand{\cteexp}{\mymath{{\vartheta}}}
\newcommand{\rrho}{\mymath{\rho_0}}

\newcommand{\shearmod}{\mymath{\mu}}   %shear modulus
\newcommand{\bulkmod}{\mymath{\kappa}} %bulk modulus
\newcommand{\poisson}{\mymath{\nu}}   %poisson ratio
\newcommand{\Cp}{\mymath{c_\mathrm{p}}}%Cp
\newcommand{\rCp}{\mymath{{}^0c_\mathrm{p}}}%Cp
\newcommand{\Cv}{\mymath{c_\mathrm{v}}}%Cv
\newcommand{\rCv}{\mymath{{}^0c_\mathrm{v}}}%Cv
\newcommand{\lat}{\boldsymbol{\ell}}%spatial latent heat
\newcommand{\cond}[1][]{\TensorTwoGreek[\Lambda][#1]}   %conductivity
\newcommand{\conds}{\mymath{\Lambda}}   %conductivity -scalar
\newcommand{\F}[1][]{\TensorTwoNonGreek[F][#1]}%c,r
\newcommand{\eF}[1][]{\mymath{\F[#1]_\elastic}}%c,p
\newcommand{\tF}[1][]{\mymath{\F[#1]_\thermal}}%t,r
\newcommand{\Fdot}[1][]{\mymath{\dot{\F[#1]}}}%c,r
\newcommand{\eFdot}[1][]{\mymath{\Fdot[#1]_\elastic}}%c,p
\newcommand{\J}{\mathrm{J}}

\newcommand{\eJ}{\mymath{\J_\elastic}}

\newcommand{\tJ}{\mymath{\J_\thermal}}

\newcommand{\bb}[1][]{\TensorTwoNonGreek[b][#1]}%c,c
\newcommand{\eb}[1][]{\mymath{\bb[#1]_\elastic}}%c,c
\newcommand{\vg}[1][]{\TensorTwoNonGreek[l][#1]}%c,r
\newcommand{\evg}[1][]{\mymath{\vg[#1]_\elastic}}

\newcommand{\Ss}[1][]{\TensorTwoNonGreek[S][#1]}%r,r
\newcommand{\Sbar}[1][]{\TensorTwoNonGreek[\bar{S}][#1]}%r,r
\newcommand{\DEVS}[1][]{\mymath{\Ss[#1]^\mathrm{dev}}}%r,r
\newcommand{\DEVSbar}[1][]{\mymath{\Sbar[#1]^\mathrm{dev}}}%r,r
\newcommand{\DEVStrial}[1][]{\mymath{{}_\trial\DEVS}}%r,r
\newcommand{\DEVSbartrial}[1][]{\mymath{{}_\trial\DEVSbar}}%r,r
\newcommand{\cauchy}[1][]{\TensorTwoGreek[\sigma][#1]}%c,c
\newcommand{\E}[1][]{\TensorTwoNonGreek[E][#1]}
\newcommand{\Edot}[1][]{\dot{\E[#1]}}
\newcommand{\eE}[1][]{\mymath{\E_\mathrm{e}}}
\newcommand{\pE}[1][]{\mymath{\E_\mathrm{p}}}
\newcommand{\tE}[1][]{\mymath{\E_\uptheta}}
\newcommand{\eEdot}[1][]{\mymath{\Edot_\mathrm{e}}}
\newcommand{\pEdot}[1][]{\mymath{\Edot_\mathrm{p}}}
\newcommand{\tEdot}[1][]{\mymath{\Edot\uptheta}}
\newcommand{\Ebar}[1][]{\bar{\E}}

\newcommand{\e}[1][]{\TensorTwoNonGreek[e][#1]}
\newcommand{\ee}[1][]{\mymath{\e_\mathrm{e}}}
\newcommand{\pe}[1][]{\mymath{\e_\mathrm{p}}}
\newcommand{\te}[1][]{\mymath{\e_\T}}
\newcommand{\ah}[1][]{\TensorTwoNonGreek[a][#1]}
\newcommand{\ahe}[1][]{\mymath{{}^\mathrm{e}\ah}}
\newcommand{\ahp}[1][]{\mymath{{}^\mathrm{p}\ah}}
\newcommand{\aht}[1][]{\mymath{{}^\T\ah}}
\newcommand{\ahs}[1][]{\mymath{{}^\mathrm{*}\ah}}

\newcommand{\Helm}{\mymath{\varphi}}
\newcommand{\eHelm}{\mymath{\Helm_\elastic}}

\newcommand{\cHelm}{\mymath{\Helm_\chemistry}}
\newcommand{\tHelm}{\mymath{\Helm_\thermal}}

\newcommand{\ien}{\mymath{e}}

\newcommand{\eW}{\mymath{W_\elastic}}
\newcommand{\eWd}{\mymath{\widehat{W}_\elastic}}

\newcommand{\eU}{\mymath{U_\elastic}}

\newcommand{\HelmDot}{\mymath{\dot{\Helm}}}

\newcommand{\x}[1][]{\vec{\TensorOneNonGreek[x][#1]}}

\newcommand{\rx}[1][]{\mymath{{}^0\x[#1]}}

\newcommand{\Body}{\mymath{\Omega}}
\newcommand{\rBody}{\mymath{\Body_0}}

\newcommand{\inv}{\mymath{^{-1}} }
\newcommand{\tr}{\mymath{^{\mathsf T}}}

\newcommand{\paren}[1]{\mymath{\!\left(#1\right)} }

\newcommand{\bracket}[1]{\mymath{\left[#1\right]} }

\renewcommand{\det}[1]{\mymath{\mathrm{det}\!\paren{#1}} }
\newcommand{\trace}[1]{\mymath{\mathrm{tr}\!\paren{#1}} }

\newcommand{\grad}[1]{\mymath{\nabla#1} }
\newcommand{\rgrad}[1]{\mymath{{}^{0}{\nabla}#1} }
\newcommand{\Div}[1]{\mymath{\nabla\cdot#1} }

\newcommand{\pderiv}[2]{\mymath{\frac{\partial #1}{\partial #2}} }
\newcommand{\spderiv}[2]{\mymath{\sfrac{\partial #1}{\partial #2}} }
\newcommand{\sppderiv}[2]{\mymath{\sfrac{\partial^2 #1}{\partial #2^2}} }

\newcommand{\ppderiv}[2]{\mymath{\frac{\partial^2 #1}{\partial {#2}^2}} }
\newcommand{\meqref}[1]{Eq.~\ref{#1}}
\newcommand{\meqrefs}[2]{Eqs.~\ref{#1}~and~\ref{#2}}

\newcommand{\figref}[1]{Figure~\ref{#1}}

\newcommand{\bs}[1]{\boldsymbol{#1}}

\newcommand{\myfigTwo}[8]{
\begin{figure}[H]
\centering
  \subfigure{
  \centering
  \label{#4}
  \includegraphics[width=#7\columnwidth]{#1}
}
\hfill
\subfigure{
  \label{#5}
  \includegraphics[width=#8\columnwidth]{#2}
}
\caption{#3\label{#6}}
\end{figure}
}

\newcommand{\order}[1]{\mathcal{O}(10^{#1})}

\begin{document}

\preprint{PRL/123-QED}

\title{Prediction of Nonlinear Specific Heat During Single Crystal HMX Phase Transition}

\author{C. W. Williams}
\author{K. Matou\v{s}}
\email{kmatous@nd.edu}
\affiliation{Department of Aerospace and Mechanical Engineering, University of Notre Dame, Notre Dame, Indiana, 46556}
\date{\today}
%%%%%%%
\begin{abstract}
We develop a thermodynamically consistent chemo-thermo-mechanical
model for the $\bd$ phase transition of energetic HMX crystals. In
contrast to previous models, which either considered specific heat to
be a constant or utilized a calibrated function, this model  provides
novel expressions for the specific heats at constant volume and
constant elastic strains derived directly from continuum mechanics. In
addition, the model provides a novel prediction for the critical
temperature at which the chemical heating rate achieves its extremum
for Arrhenius kinetics. The numerical solution predicts highly
nonlinear specific heat behavior including order of magnitude changes.
\end{abstract}
\maketitle

%%%%%%%%%%%%%%%%%%%%%%%%%%%%%%%%%%%%%%%%%%%%%%%%%%%%%
Phase transition is a chemo-thermo-mechanical (CTM) process that is
common in nature \cite{Lamberg2004, Cao2018, Enibe2003, Tan2009},
occurs in systems which are not in thermodynamic equilibrium, and is
associated with exotic material behavior.

One such exotic behavior pertains to negative specific heats known to
astronomers and \citet{LyndenBell1977} related this exotic behavior to
the large variations associated with phase transitions.

In addition to negative values in astrophysics, experiments have also
shown large changes in specific heat values for typical engineering
materials. Differential scanning calorimetry (DSC) measurements of
several phase change materials have shown order of magnitude
variations (i.e., between $\order{3}$ and $\order{4}$
[J/(kg$\cdot$K)]) \cite{Lamberg2004, Ling2013, Cao2019}. Moreover,
phase transitions in minerals have given rise to very large nonlinear
variations in their volume and density as well as softening of the
bulk modulus and other elastic constants \cite{Angel2017}. This
softening response of the bulk modulus has also been experimentally
observed during the phase transition in NIPA gels
\cite{Hirotsu1991}. It has long been recognized that conventional
phase transition models cannot successfully describe the pressure and
temperature (P-T) space, and that novel theories are required
\cite{landau1969statistical,PhysRevX.4.031010}.

In this work, we develop a novel thermodynamically consistent model to
describe the continuum level chemistry, thermodynamics, and mechanics
of materials. The model describes phase transition and makes
predictions on the exotic behavior of the specific heats. The model
also highlights the importance of the coefficient of thermal expansion
(CTE) and bulk modulus. The model is nonlinear, calibrated using
experimental data, and includes strong coupling between all relevant
fields. We also derive a novel prediction for the critical temperature
at which the chemical heating rate achieves its extremum for Arrhenius
kinetics.

We posit the plausibility of exotic specific heat behavior in phase
transitioning systems beyond those aforementioned since similar
nonlinear, non-equilibrium processes have often been observed
\cite{Baldo2002, Selbach2012, Ran2016}. In particular, we select
octahydro-1,3,5,7-tetranitro-1,3,5,7-tetrazocine (HMX), which is an
important compound for solid rocket propellants. HMX transitions
between four solid phase polymorphs \cite{Karpowicz1982}. Phase change
between the unstable $\beta$-HMX and $\delta$-HMX polymorphs is
coupled with a $6.7\%$ increase in volume \cite{Weese2005}. An
experimental effort has been directed towards understanding the
chemical \cite{Henson2002,Smilowitz2002,Weese2003,Wemhoff2007} and
thermo-mechanical \cite{Weese2005} behavior of HMX. A substantial
amount of work has also been done on the modeling and simulation of
HMX \cite{Wemhoff2007}.
%%%%%%%%%%%%%%%%%%%%%%%%%%%%%%%%%%%%%%%%%%%%%%%%%%%%%
%%%%%%%%%%%%%%%%%%%%%%%%%%%%%%%%%%%%%%%%%%%%%%%%%%%%%

We derive our model in the context of a multiplicative decomposition of the total
deformation gradient, $\F$, into thermal ($\tF$) and elastic ($\eF$)
parts given by \cite{Lee1969}
\begin{equation}
  \F(\rx,t) \equiv \rgrad{\bs{\phi}(\rx,t)} = \eF\; \tF,
\label{eq:F_split}
\end{equation}
where $\x = \bs{\phi}(\rx,t): \rBody \to \Body$ is the motion of the
body from its reference to its current configuration. Here, $\rx$ and
$\x$ are the material and spatial coordinates of the body,
respectively. We utilize an isotropic thermal deformation gradient,
\begin{equation}
  \tF = \cteexp(\T)\EyeTwo, \quad \cteexp(\T) = \exp\!\bracket{\int_{\T_0}^{\T}{\cte(\tilde\T)\diff{\tilde\T}}},
\label{eq:tF}
\end{equation}
where $\cteexp(\T)$ represents the thermal stretch ratio,
$\cte(\theta)$ is the CTE, and $\EyeTwo$ is the second order identity
tensor. For constant CTE, $\cteexp = \exp\!\bracket{\cte(\T-\T_0)}
\approx 1 + \cte (\T-\T_0)$ in the limit of small temperature
change. It is convenient to define the elastic left Cauchy-Green
tensor $\eb \equiv \eF \; \eF\tr$ which, for isotropic thermal
deformation, can be expressed as $\eb = \cteexp^{-2} \bb$ where $\bb =
\F\;\F\tr$ is the left Cauchy-Green tensor.

From here, we present the governing equations of the CTM model. The
conservation of mass reads
\begin{equation}
  \rrho = \J\rho,
\label{eq:tdf:mass}
\end{equation}
where $\rrho(\rx)$ and $\rho(\x, t)$ are the densities in the reference
and current configurations, respectively, and $\J(\rx, t) =
\det{\F}$. From the conservation of energy, we derive
\begin{align}
\label{eq:thermal:conduction}
 \rho\Cp\Tdot + \Div{\qheat} = \rho \left( \qsource + \qc  + \qe \right),
 \end{align}
where $\Cp$ is the specific heat at constant elastic strains,
$\qsource$ is the energy source per unit mass, and $\qc$ and $\qe$ are
the chemical and elastic heating terms per unit mass,
respectively. Here, $\dot{(\bullet)}$ indicates the material time
derivative. We assume Fourier's model of heat conduction, $\qheat = -
\cond \; \grad \T = -\conds \EyeTwo \grad \T$, where $\conds$ is the
isotropic thermal conductivity. The conservation of chemical species
for the $\bd$ reaction takes the form 
\begin{align}
\label{eq:chemical:conservation}
\rho \ydot_\gamma  =  \Mol_\gamma \nu_\gamma \R_c ,
 \end{align}
where $\y_\gamma$, $\Mol_\gamma$, and $\nu_\gamma$ are the mass
fraction, molar mass, and stoichiometric coefficient of the
$\gamma$-th chemical species, respectively, and $\R_c$ is the rate of
reaction \cite{Powers_Combustion_2016}. We neglect mass diffusion,
since the length scales involved are much larger than the atomic scale,
and consider $\Mol_\beta = \Mol_\delta$. Finally, the conservation of
linear momentum is
\begin{align}
\label{eq:mechanical:momentum}
\Div{\cauchy} + \rho \bforce = \vec{\boldsymbol{0}},
 \end{align}
where $\cauchy = \cauchy\tr$ is the Cauchy stress and $\bforce$ is the
body force per unit mass. Quasi-static motion is assumed since the
time scales for the problem at hand are much slower than the time
scales associated with mechanical waves. All governing equations are
solved with respect to the relevant initial and boundary conditions.

To close the system, the Helmholtz free energy reads
\begin{align}
\Helm(\T, \y_\gamma, \eb) & =  \cHelm(\T, \y_\gamma) + \tHelm(\T)  + \eHelm(\eb, \T) \label{eq:helmholtz},
 \end{align}
where $\cHelm$, $\tHelm$, and $\eHelm$, are the chemical, thermal, and
elastic parts per unit mass, respectively \cite{Vujosevic2002}. We
take $\cHelm$ to be
\begin{align}
\cHelm(\T, \y_\gamma) & = \sum_{\gamma=1}^{\Nsp}  \frac{\chempot_\gamma}{\Mol_\gamma} \y_\gamma \label{eqn:helmholtz_chemical},
\end{align} 
where $\chempot_\gamma$ is the chemical potential per mole of species
$\gamma$ which is assumed to be, at most, a linear function of
temperature. The rate of reaction, $\R_{\chemistry}$, is defined using
the law of mass action as~\cite{Powers_Combustion_2016}
\begin{equation}
\label{eq:rckc}
\R_{\chemistry} =  {\kc} \prod_{\gamma=1}^{\Nsp}  \left(\frac{\rho \y_{\gamma}}{\Mol_\gamma}\right)^{\nu^{\prime}_{\gamma}}\!\!\!,\;
{\kc} = A \dA(\Tdot)\exp{\left(-\frac{E_{a}}{R_u\T}\right)},
\end{equation} 
where we neglect the reverse reaction. Here, $\nu^{\prime}_{\gamma}$ is the
forward stoichiometric coefficient for the $\gamma$-th species. We
model the reaction constant, $\kc$, using a modified Arrhenius law
where $A$, $E_{a}$, and $R_u$ are the pre-exponential factor,
activation energy, and universal gas constant. $\dA$ is a temperature
rate dependent correction factor \cite{Wemhoff2007}.

Inspired by linear theory, which gives the canonical relation between
heat capacities $-\T( \sppderiv{\tHelm}{\T}) = \rCp = \rCv +
9\cte^2_0\T_0\bulkmod_0/\rrho$ \cite{Vujosevic2002}, we take
\begin{align}
	\label{eulerian:helmholtz_thermal}
	\tHelm (\theta) & = -\int_{\T_0}^{\T} {\int_{\T_0}^{\tilde{\T}}\left(\frac{\rCv}{\hat{\T}} + \frac{9\cte^2(\hat{\T})\bulkmod(\hat{\T})}{\rrho }\right)  \mathrm{d}\hat{\T}} \;\mathrm{d} \tilde{\T},
\end{align}
where $\rCv$ is the initial specific heat capacity at constant volume
and $\bulkmod(\T)$ is temperature dependent bulk modulus.

Finally, we take $\eHelm$ to follow the volumetric deviatoric split relation given by \cite{Simo1998,Doll1999}
\begin{subequations}
\label{eq:helmholtz_elastic}
\begin{align}
 \rrho \eHelm(\eb, \T) & = \tJ ( \eWd + \eU) = \tJ \eW, \\
\eWd(\eb, \T) & = \frac{1}{2} \shearmod(\T) [\eJ^{-2/3} \trace{\eb} - 3],\\
\eU(\eJ, \T) & = \frac{1}{4}\bulkmod(\T) \left[(\eJ-1)^2 + (\ln{\eJ})^2\right] \label{eq:helmholtz_vol},
 \end{align}
\end{subequations} 
where $\eJ = \det{\eF}$ and $\tJ = \det{\tF}$. Note the functional
dependency of $\eJ$ on $\eb$, namely $\eJ^2 = \det{\eb}$. Here,
$\shearmod(\T)$ is the temperature dependent shear modulus.

Next, we consider the Clausius-Duhem inequality~\cite{Truesdell2004},
\begin{equation}
\cauchy : \vg - \rho \HelmDot - \rho \Tdot \entropy - \frac{1}{\T} \qheat \cdot \grad \T  \geq 0, \label{eulerian:2nd_law}
\end{equation}
where $\vg = \Fdot \; \F\inv$ is the total velocity gradient. Through
the Coleman-Nole procedure \cite{Coleman-Noll}, we attain the Cauchy
stress as
\begin{align}
\label{eulerian:cauchy_eb}
\cauchy & = 2 \rho \pderiv{\Helm}{\eb}\eb = \frac{2 }{\eJ } \pderiv{\eW}{\eb}\eb,
\end{align}
with pressure, $\pressure \equiv \spderiv{\eU}{\eJ}$. We also obtain the entropy as
\begin{align}
\label{eulerian:entropy}
\entropy & \equiv \left.-\pderiv{\Helm}{\T}\right|_{\F} = \frac{\cte \cauchy : \EyeTwo \J}{\rrho} -\left.\pderiv{\Helm}{\T}\right|_{\eb}.
\end{align}
In this work, $\left.\spderiv{y}{x}\right|_{z}$ denotes the derivative
of quantity $y$ with respect to $x$ at fixed $z$. Furthermore, we
ascertain the chemical and thermal dissipation inequalities as
\begin{subequations}
\label{eulerian:Diss}
\begin{align}
\cDiss & \equiv -   \sum_{\gamma=1}^{\Nsp} \pderiv{\Helm}{\y_\gamma} \ydot_\gamma \geq 0 \label{eulerian:cDiss},\\ 
\tDiss & \equiv - \frac{1}{\T} \qheat \cdot \grad \T  \geq 0 \label{eulerian:tDiss}.
\end{align} 
\end{subequations} 
We deduce
\begin{align}
\Cp  \equiv \left. \T \pderiv{\entropy}{\T}\right|_{\eb} &= {^0}c_v+\frac{9\alpha^2\theta}{\rho_0}\bigg(\kappa+pJ\bigg) + \label{eulerian:c_def}
\\ & +\frac{3pJ}{\rho_0}\frac{\mathrm{d\alpha}}{\mathrm{d}\theta}\theta-\alpha\,\boldsymbol{\ell}_e:\boldsymbol{1}-
\theta\frac{\partial^2\varphi_e}{\partial\theta^2}\bigg|_{\boldsymbol{b}_e} \nonumber
\end{align} 
as the specific heat at constant elastic strains. Here,
$\boldsymbol{\ell}_e=-\theta(\sfrac{\partial\boldsymbol{\sigma}}{\partial\theta}\big|_{\boldsymbol{b}_e})/{\rho}$
is the spatial latent heat tensor at constant elastic strains. We note
that for temperature independent $\kappa$ and $\mu$ the specific heat
at constant elastic strains, $\Cp$, reduces to a classical specific
heat at constant pressure. The chemical and elastic heating terms are
\begin{subequations}
\label{eulerian:heating_terms}
\begin{align}
\qc & \equiv - \sum_{\gamma=1}^{\Nsp} \pderiv{\ien}{\y_\gamma}\ydot_\gamma = -{\kc} \left(1 - \y_{\delta}\right)\ereact\label{eulerian:heating_terms:qc},\\
\qe & \equiv -\bracket{2\T\pderiv{\entropy}{\eb} \eb}: \evg \label{eulerian:heating_terms:qe},
\end{align} 
\end{subequations}
where $\ien = \Helm + \T\entropy$ is the internal energy per unit
mass, $\ereact = \hreact - \Delta(3\pressure/\rho)$ is the change in
the internal energy, $\hreact$ is the heat of reaction per unit mass
and $\evg = \eFdot \;\eF\inv$ is the elastic velocity gradient. The
specific heat at constant total deformation (i.e., constant volume)
reads
\begin{align}
\Cv \equiv \left. \T \pderiv{\entropy}{\T}\right|_{\F} &=\left.-\T\ppderiv{\Helm}{\T}\right|_{\F} = \Cp -\cte  \; \lat :\EyeTwo \label{gov:eqs:cv}
\\ &=\Cp - \frac{9\alpha^2\theta}{\rho}\bigg[J_e\frac{\partial^2 U_e}{\partial J_e^2}\bigg]+\frac{3\alpha\theta}{\rho}\frac{\kappa^{\prime}}{\kappa}p,\nonumber
\end{align} 
where $ \lat \equiv -\left. { \T}  (\spderiv{\cauchy}{\T}
\right|_{\F})/{\rho}$ is the spatial latent heat tensor and
$\kappa^{\prime}=\sfrac{\mathrm{d}\kappa}{\mathrm{d}\theta}$. We note
that at the reference state, $\cte_0 = \cte(\T_0)$, $\kappa_0=\kappa(\theta_0)$,
$\eJ = 1$, $p=0$ and $\rho = \rrho$. Thus, we recover the canonical relation
between the heat capacities, where $\sfrac{\partial^2 U_e}{\partial
  J_e^2}=\sfrac{\partial p}{\partial J_e}\to \bulkmod$ for typical
volumetric potentials \cite{Doll1999}. The novel
\meqrefs{eulerian:c_def}{gov:eqs:cv} provide continuum nonlinear
descriptions of $\Cp$ and $\Cv$ for general thermo-mechanical
systems.
%%%%%%%%%%%%%%%%%%%%%%%%%%%%%%%%%%%%%%%%%%%%%%%%%%%%%
%%%%%%%%%%%%%%%%%%%%%%%%%%%%%%%%%%%%%%%%%%%%%%%%%%%%%

The model is implemented into a two-dimensional finite element solver
using a staggered isothermal split \cite{Srinivasan2009,
  Shabouei2019}. For the mechanical problem, we implement generalized
plane strain conditions \cite{Saada1983} wherein we select the motion
in the third direction such that $\cauchy_{33} \approx
\left(\cauchy_{11} + \cauchy_{22}\right)/2$, emulating an isotropic
stress response. The model is calibrated from experimental data and
calibrated parameters are within the ranges of values reported in the
literature.

%%%%%%%%%%%%%%%%%%%%%%%%%%%%%%%%%%%%%%%%%%%%%%%%%%%%%%%%%%%%%%%%%%%%%%%%%%%%%%%%
\myfigTwo{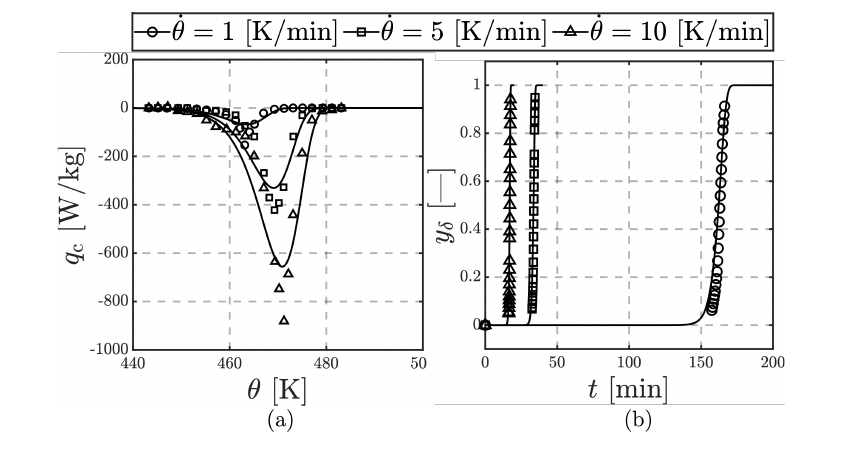}{fig1.pdf}{Calibration of the chemical model. (a)
  The chemical heating as a function of temperature, used to calibrate
  the enthalpy of reaction, $\hreact$, and the corrective factor,
  $\dA$. (b) The mass fraction as a function of time. In each graph,
  the curves are plots of the model while the markers indicate
  experimental data from
  \cite{Weese2003}.}{figs:chemistry:cal:Qc}{figs:chemistry:cal:Yd}{figs:chemistry:cal}{1.0}{0.0}
%%%%%%%%%%%%%%%%%%%%%%%%%%%%%%%%%%%%%%%%%%%%%%%%%%%%%%%%%%%%%%%%%%%%%%%%%%%%%%%%
For the chemical model, \citet{Weese2003} measured the kinetics of the
HMX $\bd$ phase transformation for heating rates of 1, 2, 5, and 10
[K/min]. They determined first order reaction parameters $A = 2.000
\times 10^{48}$ [s$^{-1}$] and $E_{a} = 432.0$ [kJ/mol]. By
substituting \meqref{eq:rckc} into \meqref{eq:chemical:conservation}
and simplifying, we find the evolution equation for the $\delta$-HMX
mass fraction
\begin{align}\label{eq:Yd:analytic}
  \ydot_\delta = \kc(1-\y_\delta).
\end{align}
%By approximating $\ydot_\delta \approx \Tdot (\spderiv{\y_\delta}{\T}) $
%and treating $\Tdot$ as fixed, this becomes an ordinary differential
%equation in $\y_\delta$ and $\T$ with the initial condition
%$\y_\delta(\T_0) = 0$. The solution reads
%\begin{align}
%    \label{eq:Yd:analytic}
%    \y_\delta(\T) & = 1 - \exp\!\paren{-A\dA\bracket{f_1(\T) + f_2(\T)}/\Tdot }, \\
%    f_1(\T) & \equiv \frac{E_a}{R_u}\mathrm{Ei}\paren{-\frac{E_a}{R_u \T}},
%    \;
%    f_2(\T) \equiv \T \exp\!\paren{-\frac{E_a}{R_u \T} }, \nonumber
%\end{align}
%where $\mathrm{Ei}$ is the exponential integral
%\cite{Gautschi1972}.
Next, we perform a least squares fit using \meqref{eq:Yd:analytic} and
\meqref{eulerian:heating_terms:qc} to the DSC heat release data
for each heating rate. The corrective factor, $\dA$, is assumed to
vary linearly as a function of temperature rate and is calibrated as
\begin{align}
\dA(\Tdot) & = 0.0451\Tdot + 0.0088. \label{eq:corrective_factor}
\end{align}
This calibration also yields enthalpy of reaction, $\hreact = 44.87$
[kJ/kg]. \figref{figs:chemistry:cal:Qc} shows the results of the
calibration for the 1, 5, and 10 [K/min] heating rates, while
\figref{figs:chemistry:cal:Yd} shows the associated $\y_\delta$
curves.

For the CTE model, \citet{Weese2005} performed measurements of the
thermal dimensional change of HMX powders. This data corresponds to
expansion with $\sim 17$\% volume change because of the CTE as well as
porosity. However, the volume change from $\beta$- to $\delta$-HMX is
6.7\% \cite{Weese2005}. With this in mind, we propose a model to
capture the overall thermal expansion as
\begin{equation}
\cte(\T) = \cte_0 + \frac{\cte_1e^{-\ctek(\T-\T_T)}}{\left[ 1+e^{-\ctek(\T-\T_T)} \right]^2 } \label{gov:eqs:cte},
\end{equation} 
where $\cte_0$, $\cte_1$, $\ctek$, and $\T_T$ are material parameters.
The axial thermal strain can be computed from
\begin{equation}
  \Delta L/L_0 = \int_{\T_0}^{\T}{\cte(\hat{\T})\diff{\hat{\T}}},
\label{eq:dL_form}
\end{equation}
where $L_0 = 3.54$ [mm] is the initial specimen length from
\cite{Weese2005}. We use the result of the integration in
\meqref{eq:dL_form} to calibrate the CTE parameters in two
stages. In the first stage, we utilize the data provided in
\cite{Weese2005} to calibrate the shape parameter, $\ctek = 0.4794$
     [K$^{-1}$], the transition temperature, $\T_T = 465.8$ [K], and
     initial guesses for $\cte_0$ and $\cte_1$. In the second stage,
     we calibrate $\cte_0$ and $\cte_1$ such that the total volume
     change, $\cteexp^3-1$, approximately matches the value of
     6.7\%. This yields $\cte_0 = 2.443 \times 10^{-5}$ [K$^{-1}$] and
     $\cte_1 = 0.007$ [K$^{-1}$]. \figref{fig:matcal:ctefits} shows
     the results of the calibration.
%%%%%%%%%%%%%%%%%%%%%%%%%%%%%%%%%%%%%%%%%%%%%%%%%%%%%%%%%%%%%%%%%%%%%%%%%%%%%%%%
\myfigTwo{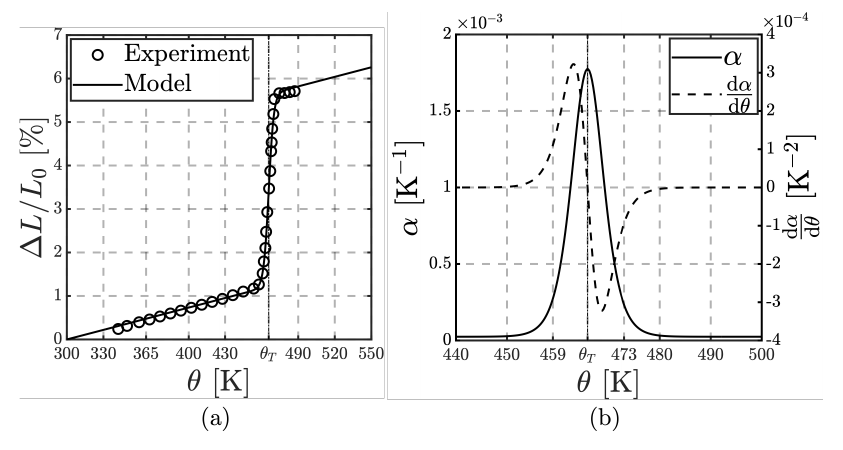}{fig2.pdf}{Calibration of the CTE model. (a)
  Thermal strain of HMX powders as a function of temperature to
  calibrate the shape parameter, $\ctek$, and CTE transition
  temperature, $\T_T$. (b) The resulting nonlinear CTE and its
  derivative. The vertical lines on each plot mark
  $\T_T$.}{fig:matcal:ctefits:a}{fig:matcal:ctefits:b}{fig:matcal:ctefits}{1.0}{0.0} 
%%%%%%%%%%%%%%%%%%%%%%%%%%%%%%%%%%%%%%%%%%%%%%%%%%%%%%%%%%%%%%%%%%%%%%%%%%%%%%%%

For the thermo-mechanical model, \citet{Dobratz1985} provide data to
calibrate the thermal conductivity as $\conds = 0.5560$
[W/(m$\cdot$K)], reference heat capacities as $\rCp = 1035$
[J/(kg$\cdot$K)] and $\rCv = 1026$ [J/(kg$\cdot$K)], and reference
density as $\rrho = 1910$ [kg/m$^3$]. Very little is known about the
behavior of the bulk modulus for HMX during the phase
transition. However, phase transition experiments on quartz
\cite{Angel2017} and on polymer gels \cite{Hirotsu1991} have shown
precipitous decrease of the bulk modulus close to the phase transition
temperature. For HMX, \citet{Levitas2004} proposed that the $\bd$
transition occurs via the stress-induced virtual melting
mechanism. Therefore, we postulate that the bulk modulus will also
substantially decrease and propose a nonlinear model given as
\begin{align}
    \bulkmod(\T) & = f_\bulkmod (\bulkmod_0 + \bulkmod_0'(\T-\T_0)) + (1-f_\bulkmod)\bulkmod_1,\\
    f_\kappa(\T) & \equiv \frac{1}{2} \bracket{1 - \tanh\paren{\frac{\xi_0}{2}}} + \frac{1}{2} \bracket{1 + \tanh\paren{\frac{\xi_1}{2}}}\nonumber,
\end{align}
where $\xi_0=\T-\T_T-\phi_0$ and $\xi_1=\T-\T_T-\phi_1$. Selection of
the material parameters $\bulkmod_0 = 11000$ [MPa],
$\bulkmod_0^{\prime}= -8.0$ [MPa/K], $\bulkmod_1 = 2000$ [MPa],
$\phi_0 = -10.8$ [K], and $\phi_1 = 9.2$ [K] are guided using
molecular dynamics simulations from \citet{Long2015} and
\citet{Cui2010}. Finally, we assume a constant Poisson ratio of
$\poisson = 0.31$ [-] \cite{Dobratz1985}, and the shear modulus is
computed using the canonical relation $\shearmod(\T) =
3\bulkmod(\T)(1-2\poisson)/(2(1 +
\poisson))$. \figref{fig:matcal:bulkmodfits} shows the calibration
results for $\bulkmod(\T)$ and $\kappa^{\prime}(\T)$.
%%%%%%%%%%%%%%%%%%%%%%%%%%%%%%%%%%%%%%%%%%%%%%%%%%%%%%%%%%%%%%%%%%%%%%%%%%%%%%%%
\myfigTwo{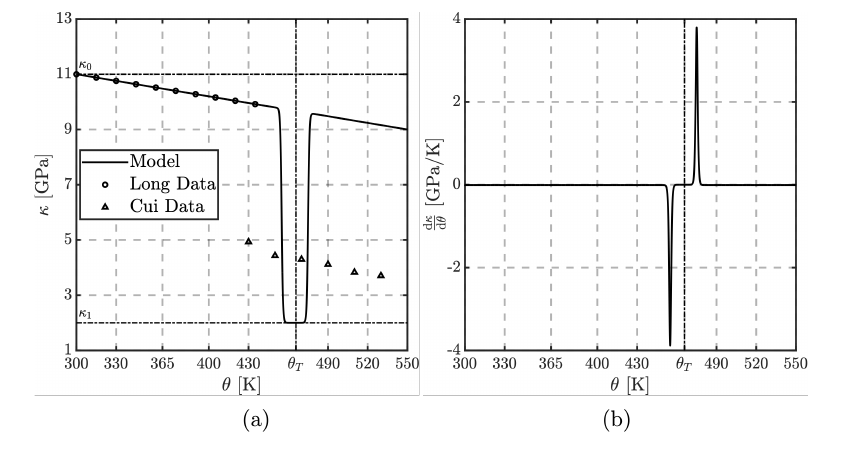}{fig3.pdf}{Calibration of the bulk modulus
  model. (a) Bulk modulus of HMX as a function of temperature. (b)
  Derivative of the bulk modulus as a function of temperature. The
  vertical lines on each plot mark
  $\T_T$.}{fig:matcal:bulkmodfits:a}{fig:matcal:bulkmodfits:b}{fig:matcal:bulkmodfits}{1.0}{0.0}
%%%%%%%%%%%%%%%%%%%%%%%%%%%%%%%%%%%%%%%%%%%%%%%%%%%%%%%%%%%%%%%%%%%%%%%%%%%%%%%%

For the numerical simulations, a $1$ mm $\times 1$ mm crystal of HMX is
heated at the boundary, $\Gamma$, at a steady rate, $\Tdot_\Gamma$, from an
initial temperature of $\T_0 = 300$ [K]. We consider temperature rates
of 1, 5, and 10 [K/min] and simulation times of 15000, 3000, and 1500
[s] to achieve a final temperature of 550 [K]. To provide well
resolved results, we have performed a mesh verification and used an
adaptive time stepping strategy as in \cite{Srinivasan2009,
  Shabouei2019}.  

\figref{figs:c} shows the specific heats at constant elastic strains,
$\Cp$, and volume, $\Cv$, averaged over the computational cell. We
observe large changes in magnitude for both specific heats (i.e.,
$\Cp$ will increase to $\approx$14840 [J/(kg$\cdot$K)] and $\Cv$ will
decrease to $\approx$528 [J/(kg$\cdot$K)]). In light of
\meqrefs{eulerian:c_def}{gov:eqs:cv}, the model predicts that the
highly nonlinear $\alpha(\theta)$, $\kappa(\theta)$, and their
derivatives (see \figref{fig:matcal:ctefits:b} and
\figref{fig:matcal:bulkmodfits}) play a large role and compete in a
highly nonlinear fashion. We note that our model predicts $\approx
6.7$\% average volume change computed as $\Delta V/V_0=\J-1$, which
compares favorably to the theoretical
estimate~\cite{Weese2005}. Furthermore, we observe a rapid temperature
rate decrease and subsequent increase due to the nonlinearity of
specific heats (see \figref{figs:dV}). However, the overall
temperature rate variations with respect to the boundary heating rate,
$\Tdot_\Gamma$, are small in part due to the crystal size.

%%%%%%%%%%%%%%%%%%%%%%%%%%%%%%%%%%%%%%%%%%%%%%%%%%%%%%%%%%%%%%%%%%%%%%%%%%%%%%%%
\myfigTwo{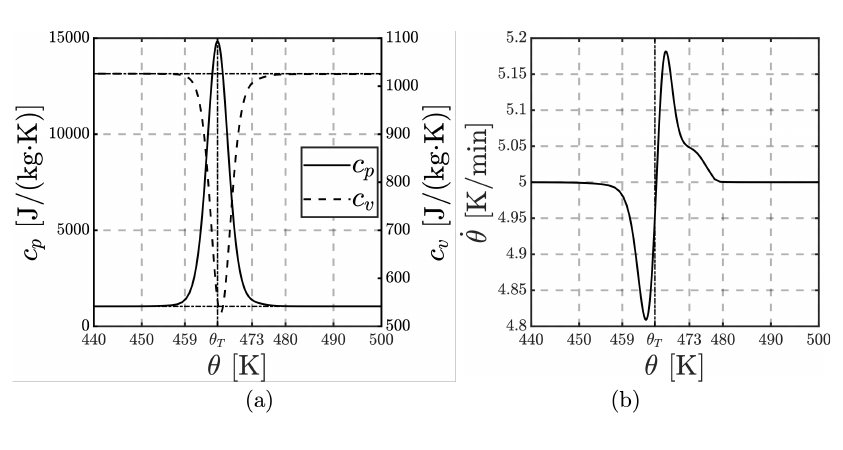}{fig4.pdf}{(a) Specific heats at constant pressure
  and volume as functions of temperature. Horizontal dotted lines
  indicate the respective reference quantities $\rCp$ and $\rCv$. (b)
  The temperature rate as a function of average cell temperature. Both
  results are from the simulation with 5 [K/min] boundary temperature
  rate.}{figs:c}{figs:dV}{figs:cdV}{1.0}{0.0}
%%%%%%%%%%%%%%%%%%%%%%%%%%%%%%%%%%%%%%%%%%%%%%%%%%%%%%%%%%%%%%%%%%%%%%%%%%%%%%%%
This is a surprising result not observed for HMX that requires a
careful analysis. First, we note that experimental results on specific
heat of HMX often consider individual phases
separately~\cite{Koshigoe1984,HansonParr1999}. Moreover, measurements
are often performed at relatively large temperature intervals
potentially under-resolving the transition that occurs over a narrow
temperature range. Furthermore, spikes in DSC traces for HMX have been
observed ~\cite{Koshigoe1984}. \citet{Levitas2004} estimated that the
elastic energy relaxed during the stress-induced virtual melting is
$\Delta h\sim 30649$ [J/kg]. Considering the transition window of
$\Delta\theta\sim 5.358$ [K] as shown in \figref{figs:c} (i.e.,
computed as an average transition temperature interval over the $\Cp$
profile), we estimate change of the specific heat during the
stress-induced virtual melting as $\Delta\Cp=\Delta h/\Delta\theta\sim
5720.2$ [J/(kg$\cdot$K)]. This value is smaller than our predictions,
but we point to large material data sensitivity of $\cte(\theta)$
and $\bulkmod(\T)$. Second, we note that nonlinear CTEs are common in
phase transitioning materials
\cite{Bolef1963,Baldo2002,Selbach2012,Ran2016} and that the $\Cp$
profile in \figref{figs:c} is similar to DSC measurements on
geopolymer concrete \cite{Cao2019}. Therefore, the nonlinearity of
$\Cp$ as predicted by our model is plausible.

Finally, we note some model limitations. Specifically, the Helmholtz
free energy, especially its thermal part in
\meqref{eulerian:helmholtz_thermal}, is not well known. Moreover, we
note the lack of pressure dependency and reaction reversibility
\cite{Karpowicz1982}, as well as crystal anisotropy, and pressure and
temperature dependency of elastic parameters, especially $\kappa$ \cite{Sewell2003}.

We continue by deriving a novel estimate for the critical temperature
at which the chemical heating rate occurs. Substituting the
$\y_\delta$ approximation, \meqref{eq:Yd:analytic}, into the chemical
heating rate, \meqref{eulerian:heating_terms:qc}, setting the
derivative of the resulting $\qc$ with respect to $\T$ equal to $0$,
then solving for $\T$, we find
\begin{align}
\theta_c (\Tdot) & \approx \frac{E_a / R_u}{2W\left( \sqrt{\left(\frac{A\dA E_a}{4\Tdot R_u}\right)}\right)} \label{eq:representative:w},    
\end{align}
where $W$ is the $W$ Lambert function \cite{Lambert1758}. For our
parameters, $\T_c$ tends to $474.2$ [K] as $\Tdot$ tends to
infinity. In \figref{figs:Qc}, we plot the chemical heating rate
averaged over the computational cell for each boundary temperature
rate. For each $\Tdot_\Gamma$, we note the associated critical
temperature, $\T_c$, at which the chemical heating rate extremum
occurs. We mark these $(\Tdot_\Gamma, \T_c)$ coordinate pairs in
\figref{figs:Qc}. In \figref{figs:Tc}, we plot the $(\Tdot_\Gamma,
\T_c)$ coordinate pairs alongside the predictions from
\meqref{eq:representative:w}. We note the remarkable agreement between
the critical temperatures from simulations and those predicted by this
equation. Furthermore, this provides a solution verification of
the computational results.
%%%%%%%%%%%%%%%%%%%%%%%%%%%%%%%%%%%%%%%%%%%%%%%%%%%%%%%%%%%%%%%%%%%%%%%%%%%%%%%%
\myfigTwo{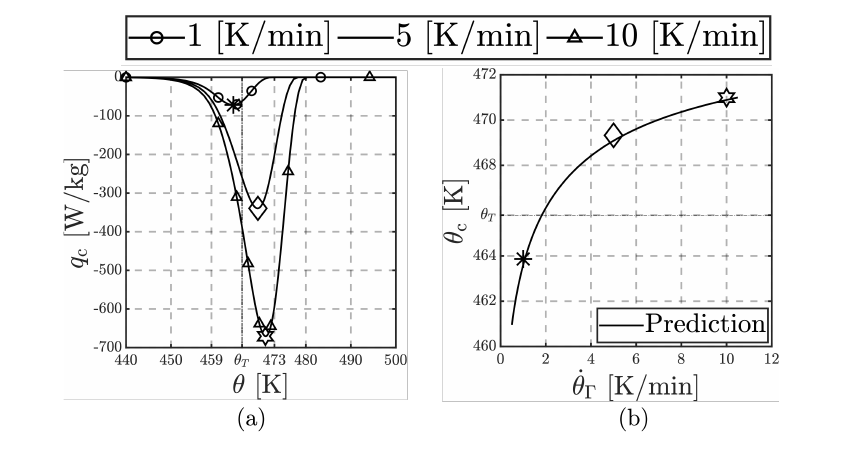}{fig5.pdf}{(a) Chemical heating rate as a function
  of temperature and (b) critical temperature for the chemical heating
  rate as a function of boundary heating
  rate.}{figs:Qc}{figs:Tc}{figs:QcTc}{1.0}{0.0} 
%%%%%%%%%%%%%%%%%%%%%%%%%%%%%%%%%%%%%%%%%%%%%%%%%%%%%%%%%%%%%%%%%%%%%%%%%%%%%%%%

In conclusion, a thermodynamically consistent continuum
chemo-thermo-mechanical model which provides general nonlinear
equations for the specific heat at constant elastic strains and volume
is derived. The model is implemented into a numerical solver and
applied to the HMX $\bd$ phase transition using parameters calibrated
with experimental data. Simulation results predict highly nonlinear,
exotic specific heat behavior including large spikes in magnitude. A
novel estimate for the critical temperature at which the chemical
heating rate extremum occurs is also derived.

%%%%%%%%%%%%%%%%%%%%%%%%%%%%%%%%%%%%%%%%%%%%%%%%%%%%%%%%%%%%%%%%%%%%%%%%%%%%%%%%
\begin{acknowledgments}
This work was supported by the Department of Energy, National Nuclear
Security Administration, under the reward No. DE-NA0002377 as part of
the Predictive Science Academic Alliance Program II. We would also
like to acknowledge support from Los Alamos National Laboratory under
award No. 625808.
\end{acknowledgments}
%%%%%%%%%%%%%%%%%%%%%%%%%%%%%%%%%%%%%%%%%%%%%%%%%%%%%%%%%%%%%%%%%%%%%%%%%%%%%%%%
%%%%%%%%%%%%%%%%%%%%%%%%%%%%%%%%%%%%%%%%%%%%%%%%%%%%%%%%%%%%%%%%%%%%%%%%%%%%%%%%
%%%%%%%%%%%%%%%%%%%%%%%%%%%%%%%%%%%%%%%%%%%%%%%%%%%%%%%%%%%%%%%%%%%%%%%%%%%%%%%%
%\bibliography{ref}
\providecommand{\noopsort}[1]{}\providecommand{\singleletter}[1]{#1}%

%%%%%%
\end{document}